\documentclass[runningheads]{llncs}
\input{psfig.sty}
\usepackage{subfigure}
\usepackage{epsfig}

\usepackage{algorithmicx}
\usepackage[ruled]{algorithm}
\usepackage[noset]{algpseudocode}
\usepackage[noset]{algpascal}
\usepackage[noset]{algc}

\newenvironment{alginc}[1][pseudocode]{\medskip\algsetlanguage{#1}\begin{algorithmic}[1]}{\end{algorithmic}\medskip}

\newcommand\ASTART{\bigskip\noindent\begin{minipage}[c]{0.5\linewidth}}

\newcommand\AENDSKIP{\end{minipage}\bigskip}
\newcommand\AEND{\end{minipage}}

\def\final{0} 

\ifnum\final=1  

\newcommand{\vnote}[1]{[{\small vicky: \bf #1}]\marginpar{*}}
\newcommand{\sidecomment}[1]{\marginpar{\tiny #1}}
\else 
\newcommand{\vnote}[1]{}
\newcommand{\nnote}[1]{}
\newcommand{\sidecomment}[1]{}

\usepackage{times}
\usepackage{amssymb}
\usepackage{fca}

\newcommand{\ms}[1]{\ensuremath{\mathsf{#1}}}

\newcommand{\union}{\cup}

\newcommand{\CC}{{\cal{B}}}
\newcommand{\OO}{{\cal{O}}} 
\newcommand{\MM}{{\cal{M}}} 
\newcommand{\II}{{\cal{I}}} 
\newcommand{\LL}{{\cal{L}}} 

\renewcommand{\int}[1]{\ms{int}(#1)}

\newcommand{\ext}[1]{\ms{ext}(#1)}
\newcommand{\nbr}[1]{\ms{nbr}(#1)}
\newcommand{\cnbr}[1]{\ms{cnbr}(#1)}
\newcommand{\UC}{\ms{Child}}
\newcommand{\IC}{\ms{AttrChild}}
\newcommand{\AC}{\ms{AttrChild}}

\newcommand{\Attr}{\ms{attr}}
\newcommand{\Obj}{\ms{obj}}

\newcommand{\res}{\ms{res}}
\newcommand{\SC}{\ms{Succ}}
\newcommand{\content}{\ms{content}}

\newcommand{\eop}{\hfill\usebox{\smallProofsym}\smallskip}  %
\newsavebox{\smallProofsym}                            
\savebox{\smallProofsym}                               %
{
\begin{picture}(6,6)                                   %
\put(0,0){\framebox(6,6){}}                            %
\end{picture}                                          %
}

\begin{document}
  \mainmatter

  \title{Faster Algorithms for Constructing
  a Concept (Galois) Lattice}
  \titlerunning{Faster Algorithms for Constructing
  a Concept (Galois) Lattice}

  \author{Vicky Choi}
  \authorrunning{Vicky Choi}
  \institute{Department of Computer Science, Virginia Tech, USA.\\
  \email{vchoi@cs.vt.edu}}

  \maketitle
  \sloppy

  \begin{abstract}
    In this paper, we present a fast algorithm for constructing a concept
    (Galois) lattice of a binary relation, including computing all concepts and their lattice
    order. We also present two efficient variants of the algorithm, one for
    computing all concepts only, and one for constructing a frequent closed
    itemset lattice.
    The running time of our algorithms
    depends on the lattice structure and is faster than all
    other existing algorithms for these problems.
  \end{abstract}

\section{Introduction}
Formal Concept Analysis
(FCA)~\cite{Ganter96Book} has found many applications since its
introduction. As the size of datasets grows, such as data generated from high-throughput
technologies in bioinformatics, there is a need for efficient algorithms for
constructing concept lattices. The input of FCA consists of a triple
$(\OO,\MM,\II)$, called {\em context}, where $\OO$ is a set of objects,
$\MM$ is a set of attributes, and $\II$ is a binary relation between $\OO$
and $\MM$. 
In FCA, the context is structured into a set of {\em
  concepts}.
The set of all concepts, when ordered by set-inclusion, satisfies the
properties of a {\em complete lattice}. 
The lattice of
all concepts is called {\em concept}~\cite{Wille82} or {\em Galois}~\cite{galois-lattice} lattice.
When the binary relation is represented as a bipartite graph, each concept
corresponds to a maximal bipartite clique (or maximal biclique). There is also a
one-one correspondence of a closed itemset \cite{Zaki98foundation} studied in data mining and a
concept in FCA. 
The one-one correspondence of all these terminologies -- concepts in FCA, 
maximal bipartite cliques in theoretical
computer science (TCS),
and closed itemsets in data
mining (DM) -- was known, e.g. \cite{Boros03inequlity,Zaki98foundation}.
There is extensive work of the related problems in these three
communities, e.g. \cite{Abello04BFS}--\cite{Tsukiyama77MIS}
in TCS, \cite{Baklouti05generalalgorithm}--\cite{Valtchev02partition}
 in FCA,
and \cite{Afrati04approximaterepresentation}--\cite{Zaki05CHARM} in DM. 
In general, 
in TCS, the
research focuses on efficiently enumerating all maximal bipartite cliques
(of a bipartite graph);
in FCA, one is interested
in the lattice structure of all concepts;  in DM, 
one is often interested in computing frequent
closed itemsets only.

\paragraph{Time complexity.} 

Given a bipartite graph, it is not difficult to see that there can be
exponentially many maximal bipartite cliques. 
For problems with potentially exponential (in the size of the input) size
output, in their seminal paper~\cite{Johnson88PolyDelay}, Johnson~et~al
introduced several notions of {\em polynomial time} for algorithms
for these problems:  {\em  polynomial total time}, 
{\em incremental polynomial time}, {\em polynomial delay time}. 
An algorithm runs in polynomial total time if the time is bounded by a
polynomial in the size of the input and the size of the output. 
An algorithm runs in incremental polynomial time if 
 the time required to generate a successive output is bounded by the size of
input and the size of output generated thus far. 
An algorithm runs in polynomial delay time if the generation
of each output is only polynomial in the size of input.
It is not difficult to see that polynomial delay is stronger than incremental
polynomial (namely an algorithm with polynomial delay time  is
also running in incremental polynomial), which is stronger than polynomial
total time.
polynomial delay algorithm, we can further
distinguish if the space used is polynomial  or exponential in
the input size.

\paragraph{Previous work.}
 Observe that the maximal bipartite 
clique (MBC) problem is a special case of the maximal clique problem in a general
graph. Namely, given a bipartite graph $G=(V_1, V_2, E)$, a maximal
bipartite clique corresponds to a maximal clique in $\breve{G} = (V_1
\union V_2, \breve{E})$ where $\breve{E} = E \union (V_1 \times V_1) \union
(V_2 \times V_2)$. Consequently, any algorithm for enumerating all maximal
cliques  in a general graph, e.g.,
\cite{Tsukiyama77MIS,Johnson88PolyDelay},  also solves the MBC problem.
 In
fact, the best known algorithm in enumerating all maximal bipartite
cliques, which was proposed by Makino and Uno~\cite{Makino04cliques} that
takes $O(\Delta^2)$ polynomial delay time where $\Delta$ is the maximum
degree of $G$,
was based on this approach. 
The fact that the set of maximal bipartite cliques constitutes a lattice was
not observed in the paper and thus the property was not utilized for
the enumeration algorithm.

In FCA, much of research has been devoted to study the properties of the lattice
structure. 
There are several algorithms,
 e.g. \cite{Nourine99basis,Valtchev02partition,Lindig00concept},
that construct the lattice, i.e. computing all concepts together with
its lattice order.
There are also some
algorithms that compute only concepts, e.g. \cite{Norris78,Ganter96Book}.
(We remark that the idea  of using a total {\em lectical order} on concepts
 Ganter's algorithm ~\cite{Ganter96Book} is also used in
 \cite{Johnson88PolyDelay,Makino04cliques}  for enumerating maximal (bi)cliques.)
See \cite{Kuznetsov02comparison} for a comparison studies of these algorithms.
The best polynomial total time algorithm was by Nourine and Raynaud
\cite{Nourine99basis}
 with
$O(nm|\CC|)$ time and $O(n|\CC|)$ space,
where $n=|\OO|$ and $m=|\MM|$ and $\CC$ denote the set of all concepts. 
This algorithm can be easily
modified to run in $O(mn)$ incremental time \cite{Nourine02incremental}. 
Observe that the space of
total size of all concepts is needed if one is to keep the entire structure
explicitly. There were several other 
algorithms, e.g.\cite{Ganter96Book,Lindig00concept}, all run in 
$O(n^2m)$ polynomial delay. 
There is another algorithm~\cite{Valtchev02partition} that is based on divide-and-conquer approach,
but the analytical running time of the algorithm is unknown as it is difficult to analyze.

There are several algorithms in data mining for computing frequent closed itemsets, such
as CHARM(-L) \cite{Zaki03Charm,Zaki05CHARM}, and
CLOSET(+)~\cite{Pei00Closet,Wang03Closetplus}. To our best knowledge, the
algorithm with theoretical analysis running time was given in 
\cite{Boros03inequlity} with $O(m^2n)$ incremental polynomial running time,
where $n=|\OO|$ and $m=|\MM|$.

\paragraph{Our Results.}
In this paper, by making use of the lattice structure of concepts, we present
a simple and fast algorithm for computing all concepts together with its
lattice order. The main idea of the algorithm is that given a concept, when
all of its successors are considered together (i.e. in a batch manner), they can be efficiently
computed. We compute concepts in the Breadth First Search (BFS) order --
the ordering given by BFS traversal of  the lattice.
When computing the concepts in this
way, not only do we compute all concepts but also we identify all
successors of each concept. 
Another idea of the algorithm is that we make use of the concepts generated
to dynamically update the adjacency relations. 
The running time of our algorithm is $O(\sum_{a \in \ext{C}} |\cnbr{a}|)$
polynomial delay for each concept $C$
(see
Section~\ref{sec:background} for related background and terminology), where
$\cnbr{a}$ is the reduced adjacency list of $a$.
Our algorithm is faster than the best known algorithms for constructing a
lattice 
because  
the algorithm is faster than a basic algorithm that runs in $O(\sum_{a \in
  \ext{C}} |\nbr{a}|)$,
 where
$|\nbr{a}|$ is number of attributes adjacent to the object $a$, and this
 basic algorithm is already
as fast as  the current best
algorithms for the problem.

We also present two variants of the algorithm: one is computing all
concepts only and another is constructing the frequent closed itemset
lattice. Both algorithms are faster than the current start-of-the-art
program for these problems.

\paragraph{Outline.}
The paper is organized as follows. In Section~\ref{sec:background}, we
review some background and notation on FCA. In Section~\ref{sec:basic}, we
describe some basic properties of concepts that we use in our lattice-construction
algorithm. 
In Section~\ref{sec:algorithms}, we first describe the high level idea of
our algorithm. Then we describe how to efficiently implement the algorithm.
In Section~\ref{sec:variants}, we describe two variants of the
algorithm. One is for computing all concepts only and another is for
constructing a frequent closed itemset lattice.
We conclude with discussion in Section~\ref{sec:discussion}.

\section{Background and Terminology on FCA}
\label{sec:background}
In FCA, a triple $(\OO, \MM, \II)$ is called a {\em context},
 where $\OO=\{g_1, g_2, \ldots, g_n\}$ is a set of $n$ elements, called {\em
 objects}; $\MM=\{1, 2, \ldots, m\}$ is a set of $m$ elements, called {\em
attributes}; and $\II \subseteq \OO \times \MM$ is a binary relation. 
The context is often represented by a {\em cross-table} as shown in
Figure~\ref{table1}. A set $X \subseteq \OO$ is called an {\em object set},
 and a set $J \subseteq \MM$ is called an {\em attribute set}.
Following the convention, we write an object set $\{a,c,e\}$ as $ace$,
 and an attribute set $\{1,3,4\}$ as $134$. 

For $i \in \MM$, denote the adjacency list of $i$ by $\nbr{i}=\{g \in \OO: (g,i)\in \II\}$.
Similarly, for $g \in \OO$, denote the adjacency list of $g$ by $\nbr{g}=\{i \in \MM: (g,i) \in \II
\}$.
\begin{definition}
  The function $\Attr : 2^{\OO} \longrightarrow 2^{\MM}$ maps a set of
  objects to their common attributes: 
  $\Attr(X) = \cap_{g \in X}\nbr{g}$, for $X \subseteq \OO$.
The function $\Obj : 2^{\MM} \longrightarrow 2^{\OO}$ maps a set of
  attributes to their common objects:
  $\Obj(J) = \cap_{j \in J}\nbr{j}$, for $J \subseteq \MM$.
\end{definition}

It is easy to check that for $X \subseteq \OO$, $X \subseteq
\Obj(\Attr(X))$, and for $J \subseteq \MM$, $J \subseteq
\Attr(\Obj(J))$.

\begin{definition}
  An object set $X \subseteq \OO$ is {\em closed} if $X =
\Obj(\Attr(X))$. An attribute set $J \subseteq \MM$ is closed if $J =
\Attr(\Obj(J))$.
\end{definition}

The composition of $\Obj$ and $\Attr$ 
 induces a {\em Galois connection} between $2^{\OO}$ and $2^{\MM}$. 
Readers are
referred to \cite{Ganter96Book} for properties of the Galois connection.

\begin{definition}
A pair $C=(A,B)$, with $A \subseteq \OO$ and $B \subseteq \MM$, is
called a
{\em concept} if $A=\Attr(B)$ and $B=\Obj(A)$.
\end{definition}

For a concept $C=(A,B)$, 
by definition, both $A$
and $B$ are closed.
The object set $A$ is called the {\em extent}  of $C$, written as $A =
\ext{C}$,  and the
attribute set $B$ is called the {\em intent} of $C$, and written as 
$B=\int{C}$.
The set of all concepts of the context
$(\OO,\MM,\II)$ is denoted by $\CC(\OO,\MM,\II)$ or simply $\CC$ when the context is
understood.

Let $(A_1,B_1)$ and $(A_2,B_2)$ be two concepts in $\CC$.
Observe that if $A_1 \subseteq A_2$, then $B_2 \subseteq B_1$. 
We order the concepts in $\CC$ by the following relation $\prec$:
$$(A_1,B_1) \prec (A_2, B_2) \Longleftrightarrow A_1 \subseteq A_2 (B_2
\subseteq B_1).$$
It is not difficult to see that the relation $\prec$ is a partial
order on $\CC$. In fact, $\LL=<\CC,\prec>$ is a
complete lattice and it is known as the {\em concept} or {\em Galois}
lattice of the context $(\OO,\MM,\II)$.
For $C, D \in \CC$ with $C \prec D$, if for all $E \in \CC$ such that $C
\prec E \prec D$ implies that $E=C$ or $E=D$, then $C$ is called the {\em
  successor} \footnote{Some authors called this as immediate successor.}(or {\em lower neighbor})
 of $D$, and $D$ is called the {\em
  predecessor} (or {\em upper neighbor}) of $C$ . The diagram representing
an ordered set (where only successors/predecessors are connected by edges)
is called a {\em Hasse diagram} (or a line diagram). See
Figure~\ref{table1} for an example of the line diagram of a Galois lattice.

For a concept $C = (\ext{C},\int{C})$, $\ext{C} = \Obj(\int{C})$ and
$\int{C}=\Attr(\ext{C})$. Thus, $C$ is uniquely determined by either its
extent, $\ext{C}$, or by its intent, $\int{C}$.
 We denote the concepts restricted to the
objects $\OO$ by 
$\CC_{\OO} = \{ \ext{C}: C \in \CC\}$, and the
attributes $\MM$ by 
$\CC_{\MM} = \{\int{C}: C \in \CC\}.$
For $A \in \CC_{\OO}$, the corresponding concept is $(A,\Attr(A))$.
For $J \in \CC_{\MM}$, the corresponding concept is $(\Obj(J), J)$.
The order
$\prec$ is completely determined by the inclusion order on $2^{\OO}$
or equivalently by the reverse inclusion order on $2^{\MM}$. 
That is,  $\LL=<\CC, \prec>$ and $\LL_{\MM}=<\CC_{\MM}, \supseteq
>$ are order-isomorphic. 
We have the property that $(\Obj(Z),Z)$ is
a successor of $(\Obj(X),X)$ in $\LL$ if and only if $Z$ is a successor of
$X$ in $\LL_{\MM}$.
Since the set of all concepts is finite, the lattice
order relation is completely determined by the covering (successor/predecessor)
relation. Thus, to construct the lattice, it is sufficient to compute
all concepts and identify all successors of each concept.

\section{Basic Properties}
\label{sec:basic}
In this section, we describe some basic properties of the concepts
on which our lattice construction algorithms are based.

\begin{proposition}
\label{Prop-1}
Let $C$ be a concept in $\CC(\OO,\MM,\II)$.
  For $i\in \MM \setminus \int{C}$, if $E_i = \ext{C} \cap \nbr{i}$ is not
  empty, $E_i$ is closed.
Consequently, $(E_i, \Attr(E_i))$ is a concept.

\end{proposition}

\begin{proof}
  For $i\in \MM \setminus \int{C}$, suppose that $E_i = \ext{C} \cap \nbr{i}$ is not
  empty. We will show that $\Obj(\Attr(E_i)) = E_i$.
Since $E_i
  \subseteq \Obj(\Attr(E_i))$, it remains to show that
  $\Obj(\Attr(E_i)) \subseteq E_i$.
By definition, $\Obj(\int{C} \union \{i\}) = (\cap_{j \in \int{C}}
  \nbr{j}) \cap \nbr{i}= \ext{C} \cap \nbr{i} = E_i$.
Thus, $(\int{C} \union \{i\}) \subseteq \Attr(\Obj(\int{C} \union
  \{i\}))=\Attr(E_i)$.
  Consequently, $\Obj(\Attr(E_i)) \subseteq \Obj(\int{C} \union \{i\})=E_i$.
\eop
\end{proof}

\begin{figure}[hbt]
\begin{center}
\begin{tabular}{c@{\quad\quad}c@{\quad\quad}c}
\hbox{
\begin{cxt}
\centering
\cxtName{}
\att{1}
\att{2}
\att{3}
\att{4}
\obj{x.x.}{a}
\obj{xx.x}{b}
\obj{x.x.}{c}
\obj{.x.x}{d}
\end{cxt}}
&\epsfig{figure=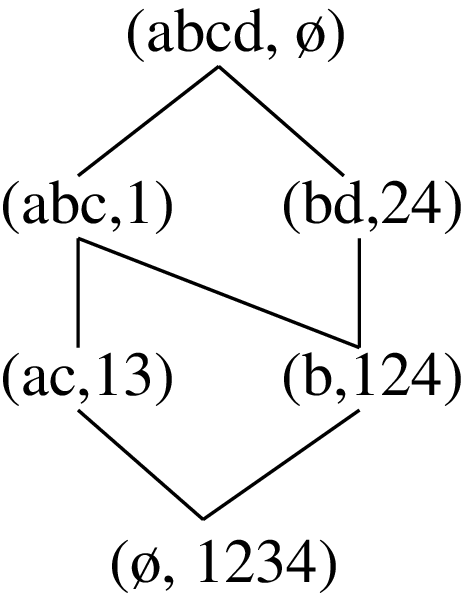,height=1in}
&\epsfig{figure=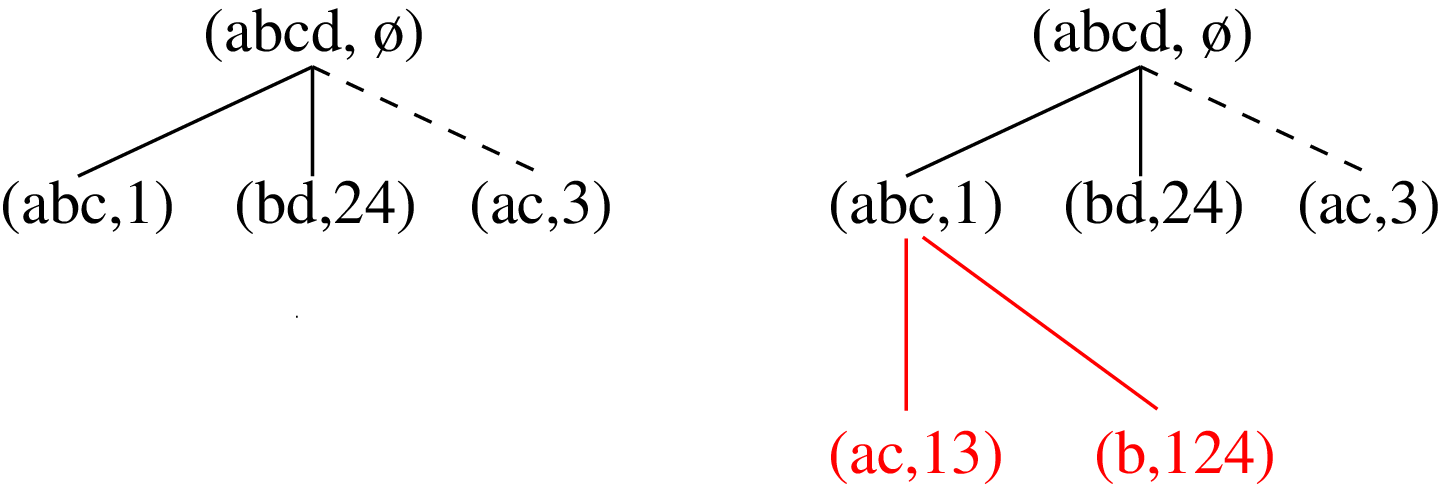,height=0.8in}\\
(a) & (b) &(c) \\
\end{tabular}
\caption{(a) 
A context $(\OO,\MM, \II)$ with $\OO=\{a,b,c,d\}$ and 
 $\MM=\{1,2,3,4\}$. The cross $\times$ indicates a pair in the
relation $\II$. (b) The corresponding Galois/concept lattice.
(c) $\UC(abcd,\emptyset)=\{(abc,1),(bd,24),(ac,3)\}$;
$\UC(abc,1)=\{(ac,13), (b,124)\}$.}
\label{table1}  
    
  \end{center}
\end{figure}

\noindent{\bf Example.}
Consider the concept $C = (abcd,\emptyset)$ of context in Figure~\ref{table1},
we have $E_1 = abc, E_2=bd, E_3=ac, E_4=bd$.

\subsection{Defining the equivalence classes}
For a closed attribute set $X \subset \MM$, 
denote the set of remaining attributes $\{i \in \MM \setminus X: 
\Obj(X) \cap \nbr{i}  \neq
\emptyset\}$ by $\res(X)$.
Consider the following equivalence relation $\sim$ on
$\res(X)$:
$i \sim j \Longleftrightarrow  \Obj(X) \cap \nbr{i}=\Obj(X) \cap \nbr{j}$,
for $i \neq j \in \res(X)$. 

Let $S_1, \ldots, S_t$ be the equivalence classes induced by
$\sim$, i.e. 
$\res(X) = S_1 \union \ldots \union S_t$, and
$\Obj(X) \cap \nbr{i}=\Obj(X) \cap \nbr{j}$ for any $i \neq j \in S_k$, $1 \leq k \leq t$.
We denote the set $\{S_1, \ldots, S_t\}$ 
by $\AC(X)$.   
We call $S_j$ the sibling of $S_i$ for $j\neq i$.
For convenience, we will write $X \union S_i$ by $XS_i$.
When there is no confusion, we abuse the notation by writing 
$X \union \AC(X) = \{XS: S \in \AC(X)\}$. 
Note that by definition, $\Obj(XS_k) = \Obj(X) \cap \Obj(S_k)=\Obj(X)
\cap \nbr{i}$ for some $i \in S_k$.
We denote the pairs $\{(\Obj(XS_1),XS_1), \ldots,(\Obj(XS_1),XS_1)\}$
by $\UC(\Obj(X),X)$. 

Recall that $\LL=<\CC, \prec>$ and $\LL_{\MM}=<\CC_{\MM},
\supseteq>$ are order-isomorphic. We have the property that $(\Obj(Y),Y)$ is
a successor of $(\Obj(X),X)$ in $\LL$ if and only if $Y$ is a successor of
$X$ in $\LL_{\MM}$.
For each $S  \in \AC(X)$, 
we call $XS$ a child of $X$ and $X$ a parent of $XS$.
By the definition of the equivalence class, 
for each $Z$ that is a  successor of $X$,   there exists a $S \in
  \IC(X)$ such that $Z=XS$. 
That is, if $Z$ is a successor of $X$, $Z$ is a child of $X$.

Let $\SC(X)$ denote all the successors of $X$, then we have $\SC(X) \subseteq X
\union \AC(X)$. 
 However, not
every child of $X$ is a successor of $X$. For the example in
Figure~\ref{table1}, $\AC(\emptyset)=\{1,24,3\}$, where $1$ and $24$ are
successors of $\emptyset$ but $3$ is not. $\SC(\emptyset) = \{1,24\}
\subset \AC(\emptyset)$; while $\AC(1)=\{24, 3\}$, $\SC(a) =\{124,13\} = 1 \union
  \AC(1)$. 
Similarly, if $P$ is a predecessor of $X$, then P is parent of $X$ but it
is not necessary that every parent of $X$ is a predecessor of $X$.

Note that for $S \in \AC(X)$, if $XS \in \SC(X)$, then by definition $XS$
is closed. 
It is easy to check that the
converse is also true. 
Namely, if $XS$ is closed, then $XS \in \SC(X)$. In other words, we have the following proposition.

\begin{proposition}
\label{succ-prop}
$\SC(X) = \{XS: XS$ is
closed, $S \in \IC(X)\}$.
\end{proposition}

\subsection{Characterizations of Closure}
By definition, an attribute set $X$ is closed if $\Obj(\Attr(X)) = X$.
In the following we give two characterizations for 
an attribute set being closed based on its relationship with its siblings.

\begin{proposition}
\label{close-prop}
For $S \in \AC(X)$, $XS$ is not closed if and only if there exists $T \in
\AC(X)$, $T \neq S$,  such that $\Obj(XS) \subset \Obj(XT)$. 
Furthermore, for all 
$T \in
\AC(X)$ with $\Obj(XS) \subset \Obj(XT)$, there exists $S' \in \AC(XT)$ 
such that $S \subseteq S'$, $\Obj(XS)=\Obj(XTS')$ and $XS \subset XTS'$.
\end{proposition}

\begin{proof}
If $XS$ is not closed, by definition, there exists $i \in
 \res(X)\setminus S$
 such that $i \in \Attr(\Obj(XS))$. As $\AC(X)$ is a partition of $\res(X)$,
 there exists a $T \in \AC(X)$ such that $i \in T$, and thus
 $\Obj(XT)=\Obj(X)\cap \nbr{i} \supset \Obj(XS)$.

Conversely,
 suppose there exists $T \in \AC(X)$ such that $\Obj(XS) \subset
 \Obj(XT)$. Then $\Attr(\Obj(XS)) \supseteq XTS$. That is, $XS \subset
 XTS \subseteq \Attr(\Obj(XS))$, which implies $XS$ is not closed.

Suppose that $\Obj(XS) \subset \Obj(XT)$ with
$T \in \AC(X)$. 
For $i\in S$, $\Obj(XT) \cap \nbr{i} = \Obj(XT) \cap \Obj(X) \cap
\nbr{i}=\Obj(XT) \cap \Obj(XS) = \Obj(XS)$. Thus,  there exists $S' \in \AC(XT)$ 
such that $S \subseteq S'$, $\Obj(XS)=\Obj(XTS')$.
Since $X,S,T$ are disjoint, $XS \subset XTS \subseteq XTS'$.
\eop
\end{proof}

Based on the first part of this proposition (first characterization), we can test if
$XS$ is closed, for $S \in \AC(X)$, by using {\em subset testing} of its
object set against its siblings' object set. 
Namely, $XS$ is closed if and only $\Obj(XS)$ is not a proper subset
of its siblings' object set. 
In our running example in Figure~\ref{table1}, $3$ is not closed because
its object set $\Obj(3) = ac$ is a proper subset of the object
set of its sibling, $\Obj(1)=abc$.

In general, subset testing operations are expensive. 
We, however, can make use of the second part of the proposition (second characterization)
for testing closure using set exact matching operations instead of
subset testing operations. 
This is because if we process the children in the decreasing order of
their object-set size, we can test the closure of $XS$ by comparing
its size against the size of the attribute set (if exists) of
$\Obj(XS)$. Namely, we first search if $\Obj(XS)$ exists by a set
exact matching operation. If it does not, then $XS$ is
closed. Otherwise, if the size of the existing attribute set of 
$\Obj(XS)$ is greater than $|XS|$, then $XS$ is not closed. 
In our running example, 
$3$ is not closed because $\Obj(3)=ac$ has a larger attribute set $13$.

\section{Algorithm: Constructing a Concept/Galois Lattice}
\label{sec:algorithms}
In this section, we first describe the algorithm in general terms, independent of the
implementation details. We then show how the algorithm can be implemented
efficiently.

\subsection{High Level Idea}
Recall that constructing a concept lattice includes generating all concepts
 and identifying each concept's successors.

Our algorithm starts with the top concept
 $(\OO, \Attr(\OO))$. We process the concept by computing all its
 {\em successors}, and  then recursively process each successor
 by either  the Depth First Search (DFS)
order --- the ordering obtained by DFS traversal of  the lattice --- or Breadth
First Search (BFS) order. 
According to Proposition~\ref{succ-prop}, successors of a concept can be
 computed from its children.  
Let $C=(\Obj(X), X)$ be a concept.
First, we compute all the  children 
$\UC(C)=\{(\Obj(XS),XS): S \in \AC(X)\}$. 
Then for each $S \in \AC(X)$,  we check if $XS$ is closed.
If $XS$ is closed, $(\Obj(XS),XS)$ is a
successor of $C$.
Since a concept can have several predecessors, it can be
generated several times. We check its existence to make
sure that each concept is
processed once and only once. 
The pseudo-code of the algorithm
based on BFS is shown in Algorithm~\ref{BFS-Lattice-alg}.

\begin{algorithm}[h]
\caption{\sc Concept-Lattice Construction -- BFS}\label{BFS-Lattice-alg}
\begin{alginc}
\State Compute the top concept $C=(\OO,\Attr(\OO))$;
\State Initialize a queue $Q =\{C\}$;
\State Compute $\UC(C)$;

\While {$Q$ is not empty}

\State $C = \mbox{ dequeue}(Q)$;
\Statex Let $X=\int{C}$ and suppose $\AC(X) = <S_1, S_2, \ldots, S_k>$;
\For {$i=1$ to $k$}
\If {$XS_i$ is closed}
\Statex Denote the concept $(\Obj(XS_i),XS_i)$ by $K$; 
\If {K does not exist}
\State Compute $\UC(K)$;
\State Enqueue $K$ to $Q$;
\EndIf
\State Identify $K$ as  a successor of $C$; 
\EndIf
\EndFor
\EndWhile
\end{alginc}
\end{algorithm}

\subsection{Implementation}
The efficiency of the algorithm depends on the efficient implementation of
processing a concept that include three procedures:
(1) computing $\UC()$; (2)testing if an attribute set is closed; (3) testing if a concept already
exists.

First, we describe  how to compute $\UC(\Obj(X),X)$ in
$O(\sum_{a \in \Obj(X)} |\nbr{a}|)$ time, using a procedure, called {\sc
  Sprout}, described in the following lemma.

\begin{lemma}
For $(\Obj(X),X) \in \CC$,
  it takes $O(\sum_{a \in \Obj(X)} |\nbr{a}|)$ to compute
  $\UC(\Obj(X), X)$.
\end{lemma}
\begin{proof}
Let $\res(X) = \union_{a \in \Obj(X)} \nbr{a} \setminus X$. 
For each $i \in \res(X)$, we associate it with a set $E_i$ (which is initialized as
an empty set).
For each object $a \in \Obj(X)$, we scan through each attribute $i$ in its
neighbor list $\nbr{a}$, append $a$ to the set $E_i$.
This step takes  $O(\sum_{a \in \Obj(X)} |\nbr{a}|)$.
Next we collect all the sets $\{E_i: i \in \res(X)\}$. 
We use a trie to
group the same object set: search $E_i$ in the trie; if not found,
insert $E_i$ into the trie with $\{i\}$ as its attribute set, otherwise we
append $i$ to $E_i$'s existing attribute set. This step takes $O(\sum_{i
  \in \res(X)}|E_i|) = O(\sum_{a \in \Obj(X)} |\nbr{a}|)$.
Thus, this procedure, called {\sc Sprout}$(\Obj(X),X)$, takes
$O(\sum_{a \in \Obj(X)} |\nbr{a}|)$ time to compute $\UC(\Obj(X),X)$.
\eop
\end{proof}

For $S \in \AC(X)$, we test if $XS$ is closed based on the second
characterization in Proposition~\ref{close-prop}.
For this method to work, it
requires  processing the children $\UC(\Obj(X),X)$ in the decreasing
order of their object-set size. 
Suppose $\AC(X)=\{S_1, \ldots, S_k\}$ where $|\Obj(XS_1)| \ge
|\Obj(XS_2)| \ge \ldots \ge |\Obj(XS_k)|$. We process $S_{i-1}$ before
$S_i$. If $XS_{i-1}$ is closed, we also compute its children
$\UC(\Obj(XS_{i-1}),XS_{i-1})$. Now to test if $XS_i$ is closed, we
check if $\Obj(XS_i)$ exists.
If it does not, then $XS_i$ is closed. Otherwise, we compare $|XS_i|$
against the size of the existing attribute set of $\Obj(XS_i)$. If $|XS_i|$
is not smaller, then $XS_i$ is closed otherwise it is not. 
To efficiently search 
$\Obj(XS_i)$, we use a trie (with hashing over each node) 
to store the object sets of concepts generated
so far and it takes linear time  to search and
insert (if not exists) an object set.
That is, 
it will take
$O(|\Obj(XS_i)|)$ time to check if $XS_i$ is closed. 
The total time it takes to check if all children are closed is
 $O(\sum_{i=1}^k |\Obj(XS_i)|)$.

Recall that a concept $C=(\Obj(X),X)$  is uniquely determined by its extent
$\Obj(X)$ or its intent $X$. Therefore, we can store either the object sets
or the attribute sets generated so far in a trie, and then test the
existence of $C$ by testing the existence of $\Obj(X)$ or $X$. Since
searching the object sets are needed in testing the closure of an attribute
set as described above, the cost of testing the existence $\Obj(X)$ comes
for free.

Note that $\sum_{a
  \in \Obj(X)} |\nbr{a}| > \sum_{i=1}^k |\Obj(XS_i)|\cdot|S_i|$. Hence, the time
  it takes to process a concept is dominated by the procedure {\sc Sprout},  in  $O(\sum_{a
  \in \Obj(X)} |\nbr{a}|)$ time.
If we can reduce the sizes of the adjacency lists
  ($|\nbr{}|$), we can reduce the running time of the algorithm.
Note that this basic algorithm is already as fast  as any existing
  algorithm for constructing a concept lattice (or computing all concepts
  only that takes $O(\Delta^2)$ time where $\Delta$ is the maximum size of
  adjacency lists).

In the following we describe how to dynamically update the adjacency lists that will
reduce the sizes of adjacent lists, and thus improve the running time of the algorithm.

\subsubsection{Further Improvement: Dynamically Update Adjacency Lists.}
\label{sec:dyn}
\begin{figure}[h]
\begin{tabular}{c@{\quad\quad}c@{\quad\quad}c}
\begin{cxt}
\centering
\cxtName{}
\att{1}
\att{2}
\att{3}
\att{4}
\att{5}
\att{6}
\att{7}
\obj{x....x.}{a}
\obj{x.xxxx.}{b}
\obj{x..x.x.}{c}
\obj{.xx.x..}{d}
\obj{.x....x}{e}
\end{cxt}
&
\epsfig{figure=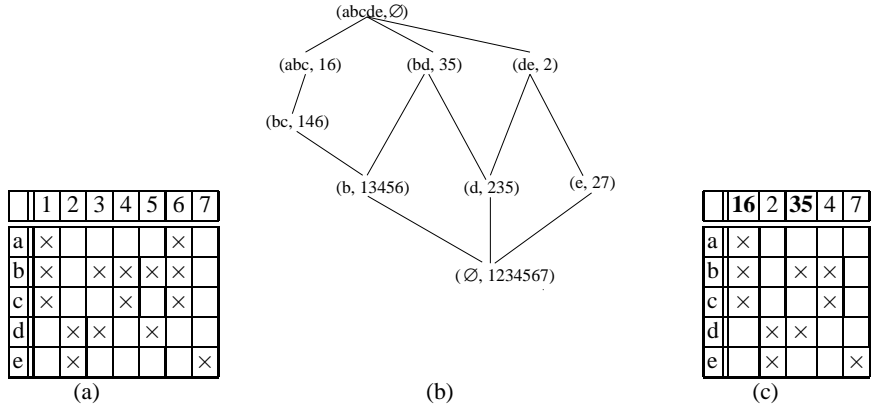,height=1.5in}
&
\begin{cxt}
\centering
\cxtName{}
\att{{\bf 16}}
\att{2}
\att{{\bf 35}}
\att{4}
\att{7}
\obj{x....}{a}
\obj{x.xx.}{b}
\obj{x..x.}{c}
\obj{.xx..}{d}
\obj{.x..x}{e}
\end{cxt}\\
(a) & (b) & (c)
\end{tabular}
  \caption{(a) A context. (b) The corresponding concept lattice. (c) Reduced cross-table of $\UC(abcde,\emptyset)$ of the context.}
\label{example2}
\end{figure}

Consider a concept $C=(\Obj(X),X)$, the object sets of all 
descendants  of $C$ are all subsets of $\Obj(X)$.
To compute the descendants of $C$, it suffices to consider
the objects with restriction to $\Obj(X)$. 
For $S \in \AC(X)$, by definition, all attributes in $S$ have the same
adjacency lists when restricting to $\Obj(X)$. That is, for all $i\neq
j \in S$, $\nbr{i} \cap \Obj(X) = \nbr{j} \cap \Obj(X) (=\Obj(XS))$.
In other words, for all $a \in \Obj(X)$, $i \in \nbr{a} \Leftrightarrow j
\in \nbr{a}$, for all $i, j \in S$, i.e., the adjacent list of $a$ either
contains all elements in $S$ or no element in $S$. Therefore, we can reduce
the sizes of adjacent lists of objects by representing all attributes in
$S$ by a single element. For example in Figure~{example2}, we can use a single
element $16$ to represent the two
attributes $1$ and $6$, and $35$ to represent $3$ and $5$. In doing so, we
reduce the size of adjacency list of $b$ from $5$ elements $\{1,3,4,5,6\}$ to
three elements $\{16,35,4\}$. 
We call
the reduced adjacency lists the condensed adjacency lists.
Denoted the condensed adjacent list by $\cnbr{}$.
The set of condensed adjacency lists corresponds to a reduced
cross-table. For example, the reduced cross table of
$\UC(abcde,\emptyset)$ of the above example
is  shown in Figure~\ref{example2}.

In order to  use the condensed adjacency lists in procedure {\sc Sprout}, 
we need to process our concepts in BFS order and it requires 
 one extra level, i.e. in a two-level manner.
More specifically, 
for a concept $C=(\Obj(X),X)$, we first compute all
its children $\UC(C)$.
Then we
dynamically update the adjacency lists by representing the attributes in each
child of $C$ with one single element.  
We then use these condensed adjacency lists to
process each child of $C$.
That is, instead of using the global adjacency
lists, when processing $(\Obj(XS),XS)$, we use the condensed adjacency
lists of its parent. 
It takes $O(\sum_{S \in
  \AC(X)}|\Obj(XS)|)$ for $C$ to generate its condensed
adjacency lists $\cnbr{}$ (see Algorithm~\ref{CondenseAdj-procedure} in the
  Appendix for the pseudo-code).
And the time for the procedure {\sc Sprout} is $O(\sum_{a \in
  \Obj(X)} |\cnbr{a}|)$ (see Algorithm~\ref{Sprout-procedure} in the
  Appendix for the pseudo-code).
Notice that $\sum_{a \in
  \Obj(X)} |\cnbr{a}| > \sum_{S \in
  \AC(X)}|\Obj(XS)|$, the time for updating the adjacency lists is
 subsumed by the time required for procedure {\sc
  Sprout}. Therefore, our new running time is 
 $O(\sum_{a \in
  \Obj(X)} |\cnbr{a}|)$ for each concept $(\Obj(X),X)$.
See Algorithm~\ref{2-level-BFS} for the pseudo-code 
and  Figure~\ref{BFStrees} for a step-by-step illustration
of the algorithm.

\section{Variants of The Algorithm}
\label{sec:variants}
For some applications, one is not interested in the entire concept
lattice. In the following, we will describe how to modify our algorithm to
solve two special cases: enumerating all concepts only and constructing a
frequent closed itemset lattice.

\subsection{Algorithm 2: Computing All Concepts or Maximal Bipartite
  Cliques}
\begin{figure}[hbt]
$$
  \begin{array}{l@{\quad\quad\quad}l}
(1)\ms{Sprout}(abcde,\emptyset):  &  (2) \ms{Sprout}(abc,16) \\
 \epsfig{figure=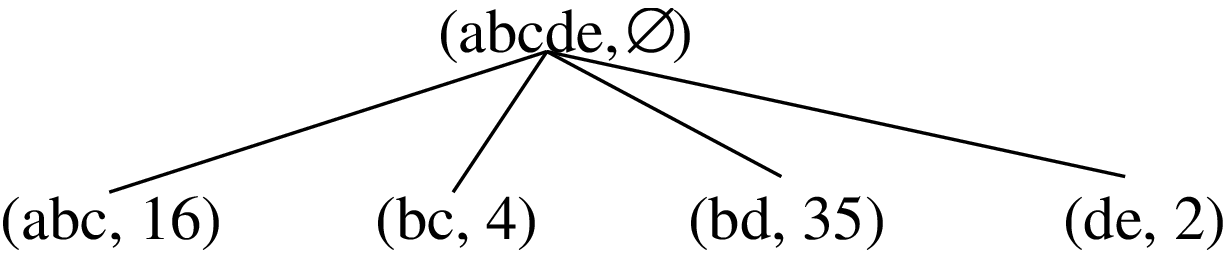,width=2.3in} &
\epsfig{figure=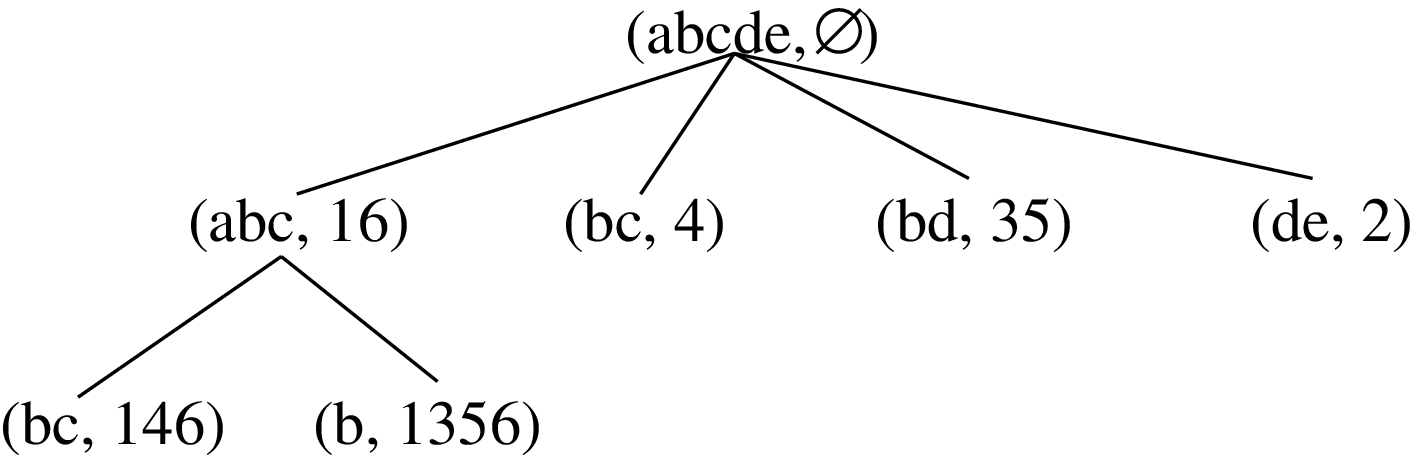,width=2.3in}\\ \\
(3) \mbox{Eliminate } (bc,4) \mbox{ as it is not closed}  & (4) \ms{Sprout}(bd,35)\\
 \epsfig{figure=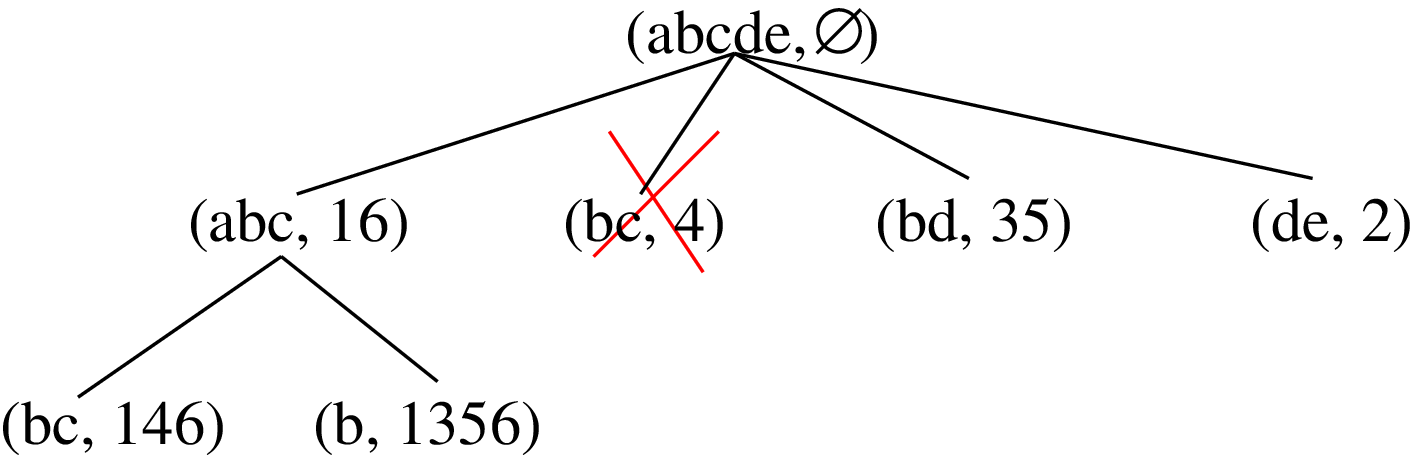,width=2.3in} &
\epsfig{figure=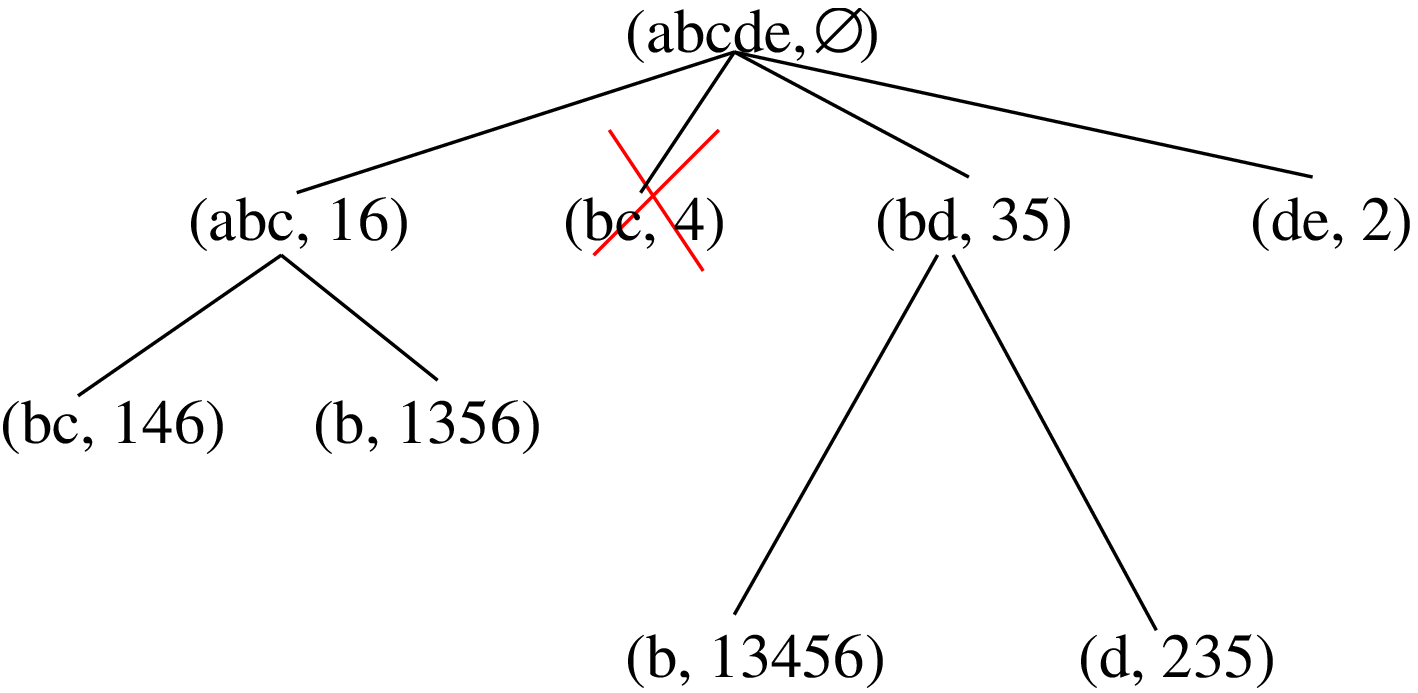,width=2.3in}\\ \\
(5) \ms{Sprout}(de,2)  & (6) \ms{Sprout}(bc,146) \\
 \epsfig{figure=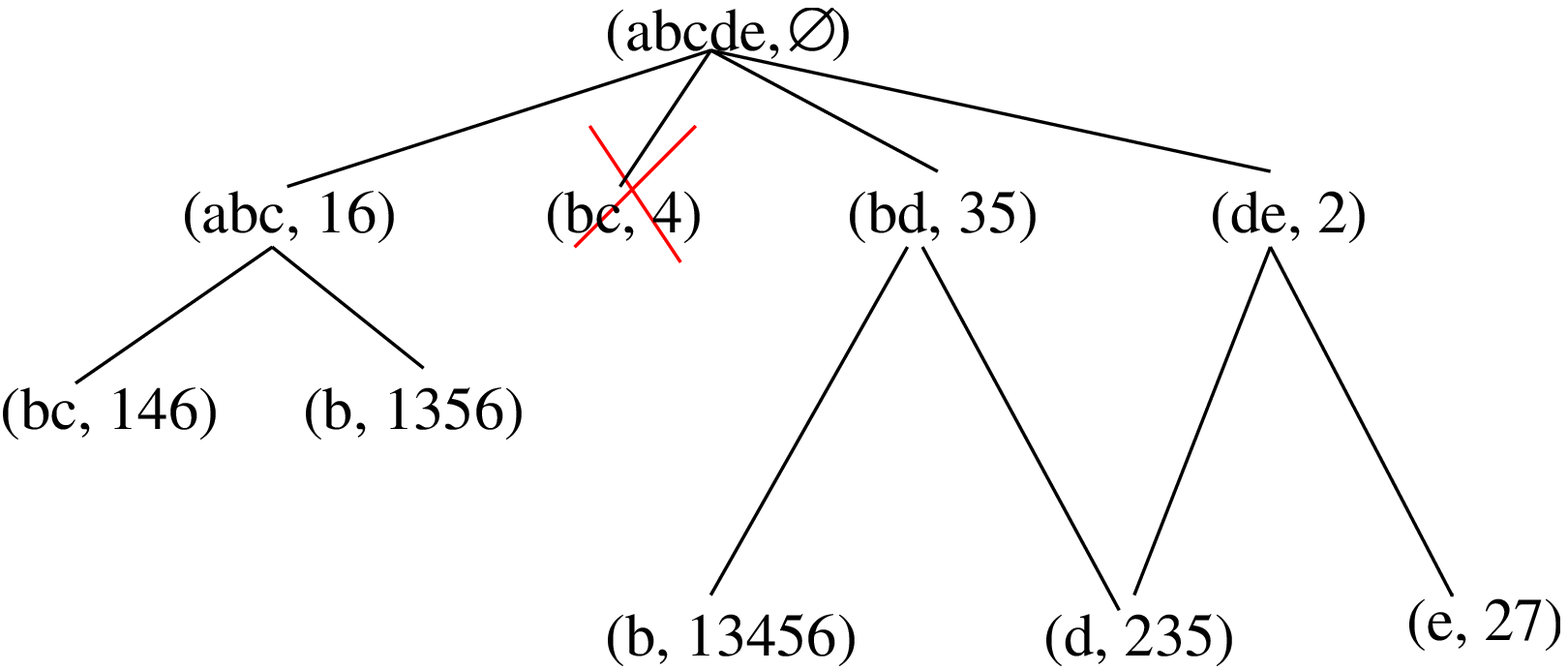,width=2.3in} &
\epsfig{figure=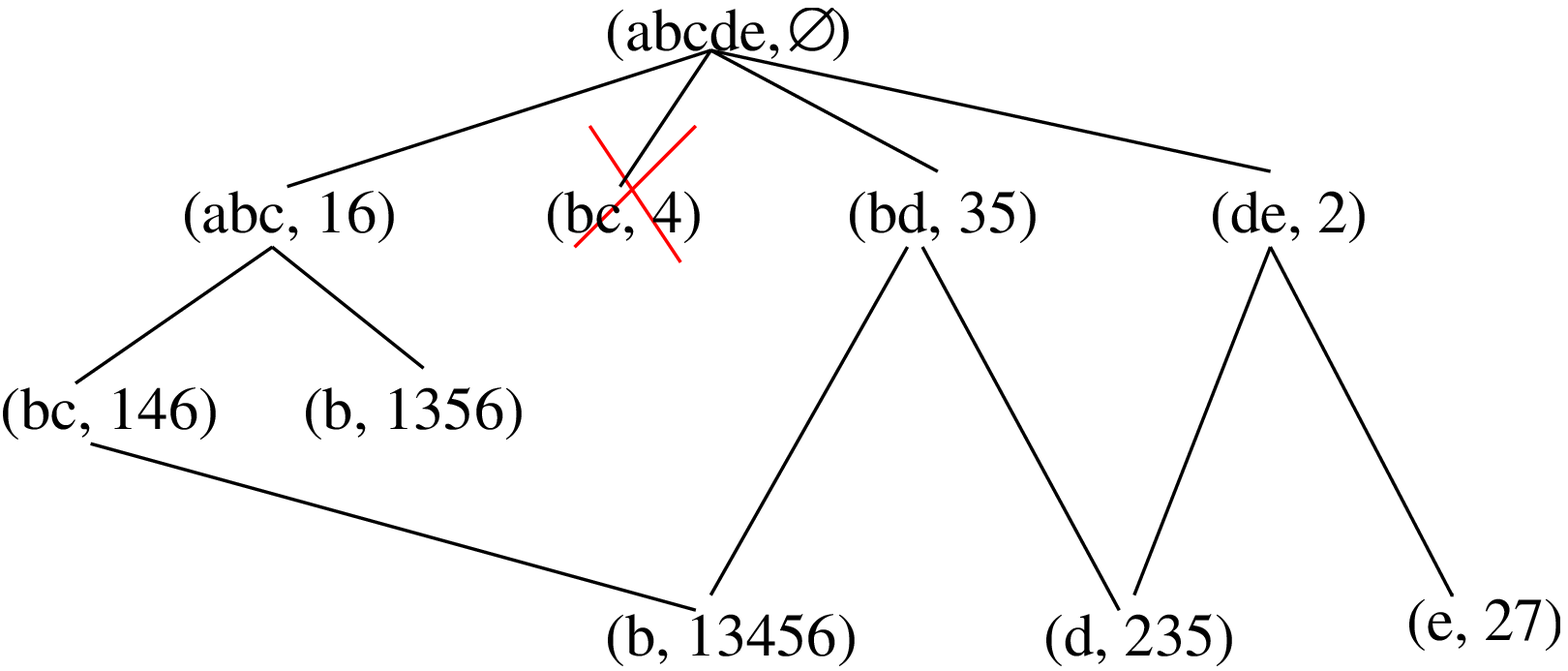,width=2.3in}\\ \\
(7) \mbox{Eliminate } (b,1356) \mbox{ as it is not closed}  &
 (8)\ms{Sprout}(b,13456), (d,235), (e, 27) \\
 \epsfig{figure=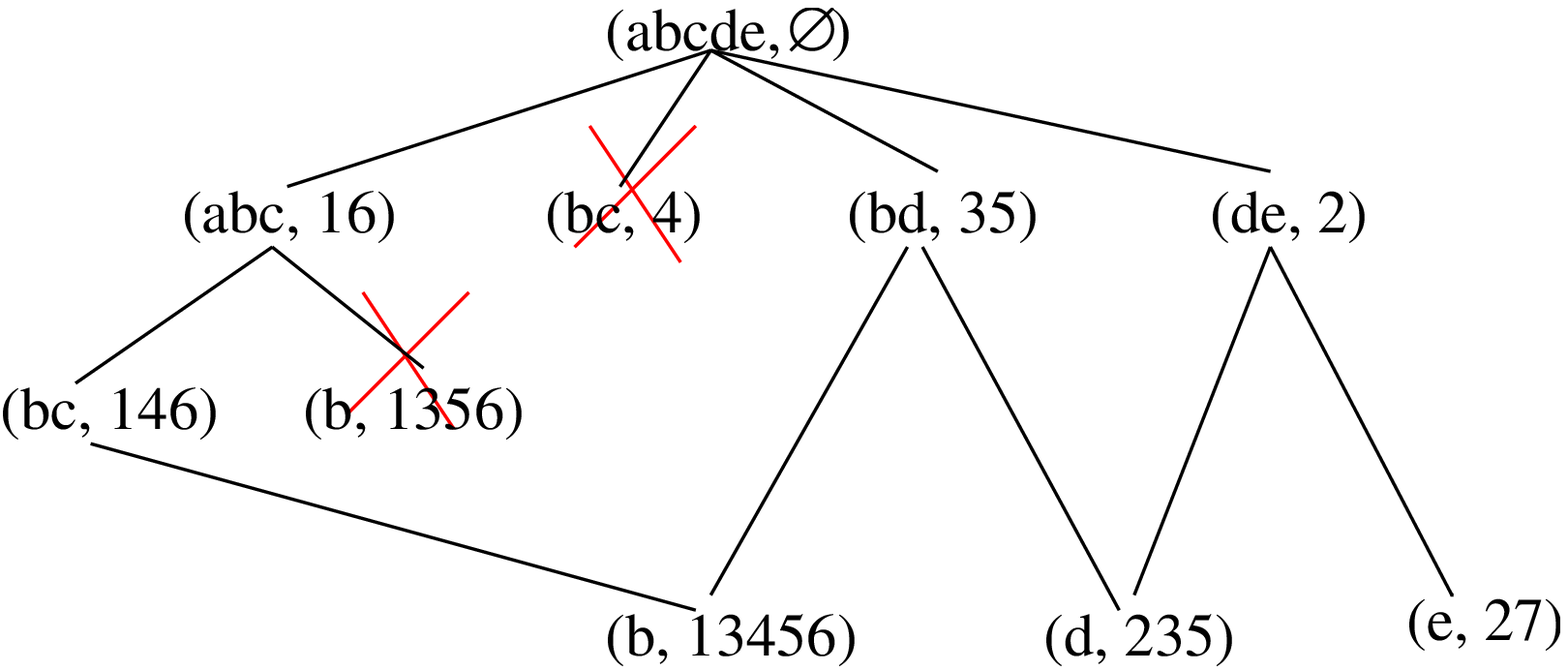,width=2.3in} &
\epsfig{figure=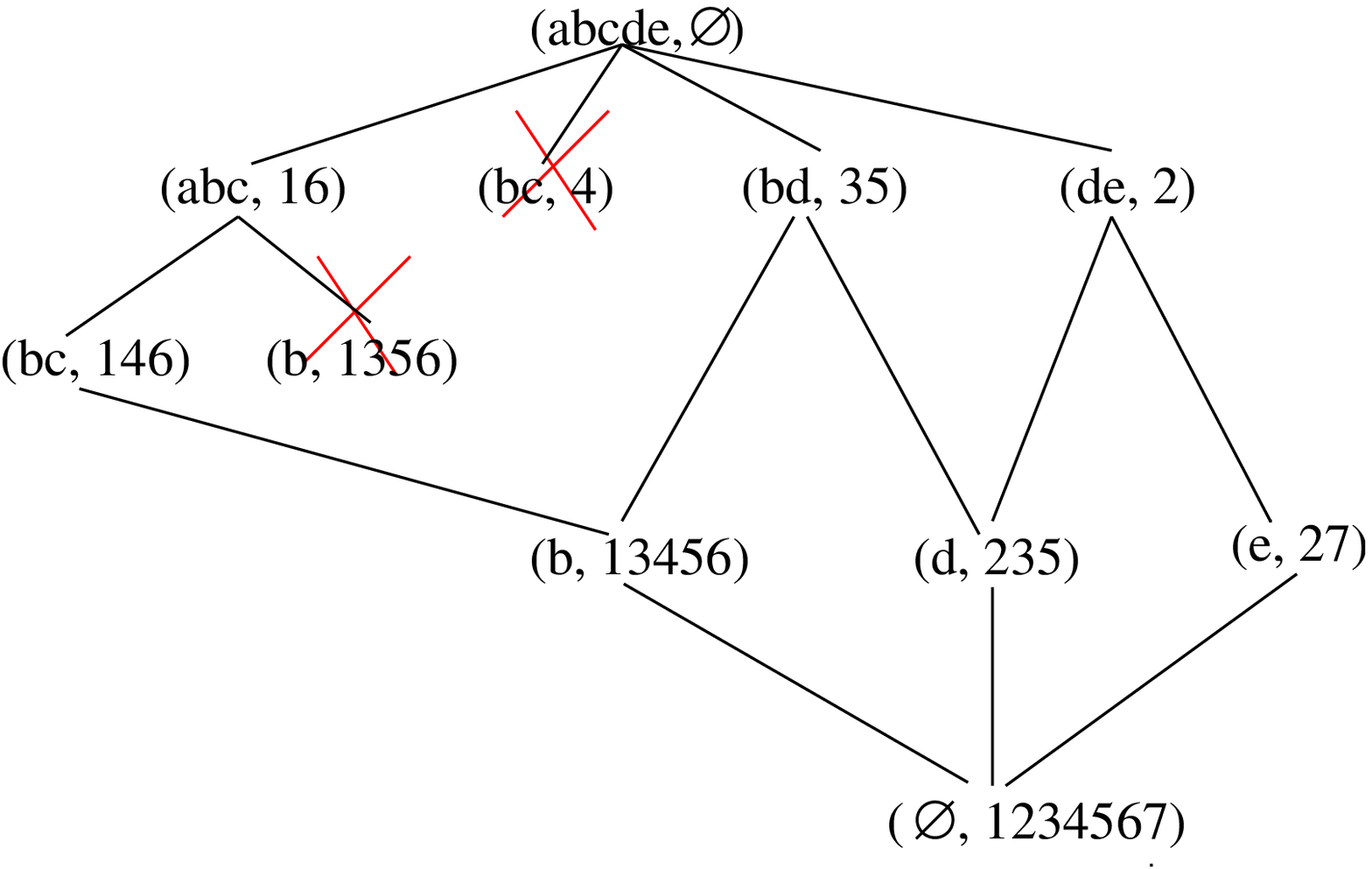,width=2.3in}\\

  \end{array}
$$
\caption{Step by step illustration of the 2-level BFS lattice construction  
 algorithm. The context and the corresponding lattice 
are shown in
  Figure~\ref{example2}. }
\label{BFStrees}  
\end{figure}

If one is interested in computing all the concepts and not in their
lattice order, as in enumerating all maximal bicliques studied in~\cite{Makino04cliques}.
We can easily modify our algorithm to give an even faster algorithm for this
purpose. This is because 
in our algorithm, each concept is generated many times, more precisely, at
least number of its predecessors times. For example in Figure~\ref{BFStrees},
$(d,235)$ is generated twice, one by each of its predecessor. 
However, when we need all concepts only, we do not
need regenerate the concepts again and again. This can be easily
accomplished by considering the right siblings only in the procedure {\sc
  Sprout}, i.e. changing the line 3 to $\ms{for}~~i \in \nbr{a} \mbox {
  AND } i>s~~\ms{do}$, while the other parts of the algorithm remain the
same. 
Depending on the lattice structure, this can significantly speed up the
algorithm as the number of siblings is decreasing in a cascading
fashion. A more careful analysis is
needed for the running time of this algorithm.

\subsection{Algorithm 3: Constructing a Closed Itemset Lattice}
In data mining, one is interested in large concepts,
i.e. $(\Obj(X),X)$ where $|\Obj(X)|$ is larger than a threshold. 
Although
our algorithm can naturally be modified to construct such a closed itemset
lattice:  we stop processing a concept when the size of its object set is less
than the given threshold, where
objects correspond to  transactions and attributes
correspond to items. Theoretically, when the memory requirement is not
a concern, our algorithm is faster than all other existing algorithms
(including the state-of-art program CHARM-L) for
constructing such a frequent closed itemset lattice.
 However,
in practice, for large data sets (as those studied in data mining), the
data structure -- a trie on objects (transactions) -- requires huge memory
and this may threaten the algorithm's practical efficiency. However, it
is not difficult to modify our algorithm so that a trie on attributes
(items) instead is used. Recall that a trie on objects are required in two
steps of our algorithm: testing the closure of an attribute set and testing
the existence of a concept. As noted above, the existence of a concept can
also be tested on its intent (i.e. attributes), thus we can use a trie on
attributes 
for testing the existence of a
concept. To avoid using a trie on objects for testing the closure of an
attribute set, we can use the first characterization in Proposition~\ref{close-prop} instead,
that is, we test the closure of an attribute set 
by using {\em subset testing} of its
object set against its siblings' object set, as described in Section~\ref{sec:basic}.
Further,
we can employ the practically efficient technique {\em
  diffset} as in CHARM(-L) for both our {\sc Sprout} procedure and subset
testing operations. We are testing the performance of the diffset based
implementation on the available benchmarks and the results will be reported
elsewhere. 

\vspace*{-0.5cm}

\section{Discussion}
\label{sec:discussion}
Our interest in FCA stems from our research in microarray data analysis \cite{V-DIMACS}. We
have implemented an not yet optimized version of our algorithm (with less than
500 effective lines in C++). The program is very efficient for our
applications, in which our data consists of about 10000 objects and 29
attributes. It took less than 1 second for the program to produce the
concept lattice (about 530 vertices/concepts and 1500 edges) in a Pentium
IV 3.0GHz computer with 2G memory running under Fedora 2 linux OS. The program is available upon
request at this point and will release to the public in the near future.

As FCA finds more and more applications, especially in 
 bioinformatics, efficient algorithms for constructing concept/Galois lattices are
 much needed. Our algorithm is faster than the existing algorithms for
 this problem, nevertheless, it seems to have much room to improve.

 \vspace*{-0.5cm}
\section*{Acknowledgment}
We would like to thank Reinhard Laubenbacher for introducing us FCA. We
thank Yang Huang for his participation in his project.

\newpage
\begin{center}
  Appendix
\end{center}

\begin{algorithm}[h]
\caption{{\sc Sprout}}\label{Sprout-procedure}
\begin{alginc}
\Statex Input: $s$, $\content$ and $\ms{nbr}$ 
\Statex $(\Obj(X), X)$ is the $s$th child of $G$. Let $K = \{1,\ldots, k\}$ be all the
children of $G$.

\Statex Output: $\UC(\Obj(X),X) = \{(\Obj(XS_i), XS_i): 1 \leq i \leq
t\}$

\State For each $i \in K$, set $C_i =
\emptyset$.
\For {$a \in C$}
\For {$i \in \nbr{a} \setminus \{s\}$} 
\State Append $a$ to $C_i$;
\EndFor
\EndFor

\Statex The following takes $O(\sum_{i \in K}
|C_i|) = O(\sum_{a \in C}|\nbr{a}|)$ time.

\State Initialize a local trie $T_C$ over objects;
\For {$i \in K$}
\If {$C_i$ does not exist in $T_C$}
\State Insert $C_i$ into $T_C$; 
\State $S_i = \content(i)$;
\Else
\State Merge $S_i$ with $\content(i)$ ;
\EndIf
\EndFor
\State Output all the pairs in $T_C$: $\{(\Obj(XS_j), XS_j): 1 \leq j
\leq t\}$.
\end{alginc}
\end{algorithm}

\begin{algorithm}[h]
\caption{{\sc CondenseAdjacentlists}}\label{CondenseAdj-procedure}
\begin{alginc}
\Statex Input: $\UC(C)=\{(\Obj(XS_i), XS_i): 1 \leq i
\leq t\}$ 

\Statex Output: $\content(i)$ for $i=1 \ldots t$, and new adjacency lists, $\ms{nbr}(a)$,  $a \in \Obj(X)$

\State For each $a \in \Obj(X)$, $\nbr{a}=\emptyset$;
\For {$i=1$ to $t$}
\State $\content(i)=S_i$;
\State for each $a \in \Obj(XS_i)$, append $i$ to $\nbr{a}$;
\EndFor
\end{alginc}
\end{algorithm}

\begin{algorithm}[h]
\caption{\sc Concept-Lattice Construction -- 2-level BFS}\label{2-level-BFS}
\begin{alginc}
\State Compute the top concept $C=(\OO,\Attr(\OO))$;
\State Initialize a queue $Q =\{C\}$;
\State Initialize a trie $T$ for the object set $ \OO$;

\State $\content(i)=\{i\}$ for $i \in \MM$;
\State $\ms{Child(C)}=$ {\sc Sprout}$(0, \content, \ms{nbr})$;

\While {$Q$ is not empty}

\State $C = \mbox{ dequeue}(Q)$;
\State Sort the pairs in $\ms{Child}(C)$ according to its extent
size in decreasing order: $(\Obj(XS_i), XS_i), 1 \leq i \leq k$.

\State $(\content,\ms{nbr})=$ {\sc CondenseAdjacentlists}($\ms{Child}(C)$);

\For {$i=1$ to $k$}
\State Search $\Obj(XS_i)$ in $T$; 
\Statex Denote $(\Obj(XS_i),XS_i)$ by $K$;
\Comment $K$ is not necessary a concept.
\If {$\Obj(XS_i)$ does not exist}
\State Insert $\Obj(XS_i)$ into $T$, and associate it with the attribute
set $XS_i$;
\State Identify $K$ as the successor of $C$;
\State $\ms{Child}(K)=$ {\sc Sprout}$(i,\content,\ms{nbr})$;
\State Enqueue $K$ into $Q$;
\ElsIf {the attribute set associate with $\Obj(XS_i)$ is not greater than
  $XS_i$}
\State Identify $K$ as the successor of $C$;
\EndIf
\EndFor
\EndWhile
\end{alginc}
\end{algorithm}

\end{document}